\documentclass[12pt]{spieman}  
\usepackage{amsmath,amsfonts,amssymb}
\usepackage{graphicx}
\usepackage{setspace}
\usepackage{tocloft}
\usepackage{xcolor}

\title{Data post-processing gain resulting from the patchy nature of speckles}

\author[a,*]{Jean-Baptiste Ruffio}
\author[b]{Laurent Pueyo}
\affil[a]{Department of Astronomy \& Astrophysics, University of California San Diego, La Jolla, CA, USA}
\affil[b]{Space Telescope Science Institute, 3700 San Martin Drive, Baltimore, MD 21218, USA}

\cftpagenumbersoff{figure}
\cftpagenumbersoff{table} 
\begin{document} 
\maketitle

\begin{abstract}

The data post-processing gain is an important parameter for exposure time calculations used to inform the design of the Habitable Worlds Observatory (HWO). 
Assuming azimuthally symmetric noise properties is a common simplifying assumption for such simulations, which neglects the patchy nature of the residual diffracted starlight; i.e., speckles. Fortunately, patchiness might prove to be an opportunity that improves the overall sensitivity of observatory assuming photon-noise limited speckle subtraction. 
We illustrate this effect in the context of angular differential imaging (ADI), which is one of the possible observing strategies being considered for the detection and characterization of exo-Earth with HWO.
We show that combining observations of two observatory roll angles leads to a gain in signal-to-noise greater than $\sqrt{2}$ when the patchy starlight dominates other noise sources. 
The gain can be closer to $\times2$ when the starlight dominates the noise budget by more than an order of magnitude.
In other words, combining good and bad observations is better than combining two average ones. This statement is very general as it is a direct consequence of combining data with a weighted mean. It applies more broadly to any combination of observations with varying noise level.
\end{abstract}


{\noindent \footnotesize\textbf{*}Jean-Baptiste Ruffio,  \linkable{jruffio@ucsd.edu} }

\begin{spacing}{2}   

\section{Introduction}
\label{sect:intro}  

Direct imaging of exoplanet relies on methods to separate the planet signal from the diffracted starlight; the speckles \cite{Currie2023ASPC..534..799C}. Speckles are by definition ``patchy''; without post-processing, they appear as bright copies of the instrument off-axis response separated by dark regions. This patchiness remains in the form of variations of the shot noise level after point spread function (PSF) subtraction, even with perfect calibration of the systematics. For example, a larger photon noise level will be found around the position of a bright speckle. This means that the photon noise affecting the planet signal depends on the chance alignment of the planet relative to the features of the speckle pattern. The signal-to-noise ratio (S/N) of a planet should therefore depend, not only on the angular separation relative to its host star, but also on the azimuth angle. 
The latter is often neglected in practice. For example, the standard deviation of the noise is often estimated empirically in concentric annuli around the star as a function of projected separation\cite{Marois2006ApJ...641..556M}. However, obtaining accurate 2D sensitivity maps that account for the 2D nature of the speckles pattern is possible as shown with JWST/NIRSpec IFU\cite{Ruffio2024AJ....168...73R}. 
Creating 2D noise models is necessary to more optimally combine observations.
The goal of this work is to highlight the effect of the non-uniformity of the speckle noise on the achieved planet sensitivity when 2D noise models are available. We will use angular differential imaging (ADI)\cite{Marois2006ApJ...641..556M} as an illustration, but the results are not limited to ADI. In the context of space-based observatories, ADI consists in rotating the telescope to obtain at least a pair of exposures in which the speckle pattern is rotated compared to the planet.

The effect of speckle non-uniformity is relevant for exoplanet yield calculation used to inform and optimize the design of the Habitable World Observatory (HWO)\cite{Stark2019JATIS...5b4009S,Stark2024JATIS..10c4006S}. Yield calculations rely on exposure time calculators (ETC) to estimate the exposure time, or S/N, of a planet given a set of assumptions and observing strategy. One parameter of such simulations is the post-processing gain that factors in any improvement in sensitivity due to post-processing algorithms compared to the raw starlight level provided by the starlight suppression system\cite{Nemati2020JATIS...6c9002N}. Such simulations are typically assumed to be azimuthally symmetric, meaning that the sensitivity is only a function of the projected separation. For starlight-dominated systems, we show below that the patchiness of the starlight can be approximated by an additional data post-processing gain.

Using the concept of weighted mean, we aim to answer the following question: how does the predicted S/N change when including knowledge about the speckle intensity variations compared to azimuthally-averaged noise assumptions? Leveraging the non-uniformity of the speckles is however only possible if accurate 2D noise maps are available.

\section{Combining two observations with a weighted mean}

In this section, we demonstrate that combining observations from two rolls lead to an improvement of the S/N greater than $\sqrt{2}$. We assume that the planet lands respectively on a bright and a faint spot of the speckle pattern around the average starlight level.
In practice, this can be achieved by ensuring that the observatory roll angle results in a displacement of the planet relative to the speckles that is similar to the spatial resolution ($\sim\lambda/D$) of the instrument. This rotation will however only be valid for a limited range of projected separations.
We assume an idealized photon noise-limited speckle subtraction. This means that we assume the existence of a noise-free model of the stellar PSF without systematics. We note that if an empirical PSF is used, e.g. for pair-wise PSF subtraction, then the background and starlight fluxes should be doubled to account for the increased noise due to the subtraction. Inaccuracies in the noise model, for example when it is estimated empirically from the data, could lead to biases for planet detection. While it would be important to address in practice, this particular issue is outside the scope of this work.

\begin{figure}
\begin{center}
\begin{tabular}{c}
\includegraphics[height=7cm]{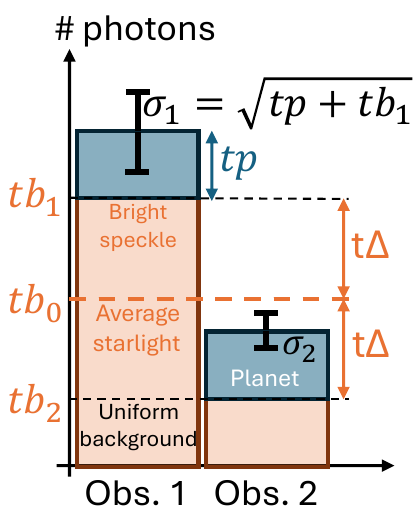}
\end{tabular}
\end{center}
\caption 
{ \label{fig:goodbad}
Illustration of two observations of a planet with different underlying starlight noise levels. The starlight and other background sources are represented in orange, while the planet signal is shown in blue. In observation 1, the planet sits on top of a bright speckle. In observation 2, the speckle flux is null and the planet only sits on top of the uniform background light. The mathematical notations are defined in the text.} 
\end{figure} 

We define two observations as illustrated in Fig.~\ref{fig:goodbad}. Observation 1 assumes that the planet landed on top of a bright speckle and observation 2 has the planet in a speckle free region. We define the following variable:

\begin{itemize}
    \item $t$ is the exposure time of a single roll. The total exposure time for two rolls is therefore $t_\mathrm{tot}=2t$.
    \item $p$ is the photon count rate for the planet signal. The number of planet photons is $tp$.
    \item $b_0$ is defined as the average background photon count rate including the average starlight level. 
    \item $b_1$ is the peak background photon count rate corresponding to a bright speckle in observation 1. It includes any uniform background component as well.
    \item $b_2$ is the uniform background light contribution in observation 2, which might include the exozodi, zodi, or detector noise.
    \item $\Delta$ is the average starlight-only component of the background such that $b_1 = b_0 + \Delta$ and $b_2 = b_0 - \Delta$. 
    \item $\sigma_i^2 = t( p + b_i)$ is the variance of the Poisson noise in each observation assuming photon noise-limited observations.
\end{itemize}
The signal-to-noise ratio (S/N) in each observation is therefore defined as $S/N = tp/\sigma_i$.

Combining two observations with different noise level can be accomplished with a weighted mean. The combined expected estimate of the planet photon count for a single roll of time $t$ is still $\tilde{tp}=tp$ while the estimated combined variance of the noise is given by:
\begin{align}
    \tilde{\sigma}^2 &= \frac{1}{\frac{1}{\sigma_1^2} + \frac{1}{\sigma_2^2}}, \nonumber \\
   \tilde{\sigma}^2 &= t\times\frac{p + b_0}{2} \left(1 - \left(\frac{\Delta}{p + b_0}\right)^2\right).
\end{align}

This leads to the following combined S/N:
\begin{align}
    \widetilde{S/N} &= \frac{t\tilde{p}}{\tilde{\sigma}}, \nonumber \\
    &= \widetilde{S/N}_0 \left(1 - \left(\frac{\Delta}{p + b_0}\right)^2\right)^{-1/2},
    \label{eq:snr}
\end{align}
with $\widetilde{S/N}_0 = \frac{\sqrt{t_\mathrm{tot}}\times p}{\sqrt{p + b_0}}$, which is the classical combined S/N when both observations have equal noise (i.e., $\Delta=0$) and the S/N improves with square root of time.

\textbf{We therefore define the additional post-processing gain compared to the uniform noise assumption as:}
\begin{equation}
    g =  \left(1 - \left(\frac{\Delta}{p + b_0}\right)^2\right)^{-1/2}
     \label{eq:gain}
\end{equation}

For example, $g\approx1.15$ if $p=b_0=\Delta$, which means that the planet is as bright as the average starlight and there is no uniform component of the background noise. The gain becomes more important when the planet is significantly fainter than the starlight; indeed, $g\approx2.4$ if $10\times p=b_0=\Delta$.

For a given observation, the value of $\Delta$ will depend on the random chance alignment of the planet signal with speckles, which will reduce the post-processing gain because the effective $\Delta$ on average is smaller. This is the topic of the following sections.

\section{Average processing gain: assuming infinite rolls}
\label{sec:avggain_inf}

In this section, we aim to compute the average processing gain in the case of an infinite number of rolls. We will do so by first assuming $N_\theta$ rolls and then estimating the limit when $N_\theta\rightarrow\infty$. 
The inverse combined variance is given by:

\begin{equation}
\frac{1}{\tilde{\sigma}^2} = \frac{1}{\sigma_1^2} + \frac{1}{\sigma_2^2} + \frac{1}{\sigma_3^2} + \cdots,
\end{equation}
with $\sigma_i^2 = t_{N_\theta} (p+ b(\theta_i))$ the variance at each rotation angle $\theta_i$ and $t_{N_\theta}=t_\mathrm{tot}/N_\theta$ is the time of a single roll.

Following the same principles as Eq.~\ref{eq:snr}, the combined S/N is now:

\begin{align}
    \widetilde{S/N}  &= \frac{t_{N_\theta}\tilde{p}}{\tilde{\sigma}}, \nonumber \\
    &= \widetilde{S/N}_0 \frac{1}{\sqrt{N_\theta}}\frac{\sigma_0}{\tilde{\sigma}}, \nonumber \\
    &= \widetilde{S/N}_0 \left(  \sigma_0^2 \times \frac{1}{N_\theta} \sum_{i=1}^{N_\theta} \frac{1}{\sigma_i^2}  \right)^{1/2},
    \label{eq:snr}
\end{align}

Then, we note that by definition $ \sigma_0^2 = \langle \sigma_\theta^2 \rangle_\theta$, so at the limit of an infinite number of roles, 
the S/N gain over $\widetilde{S/N}_0$ becomes

\begin{equation}
    g_\infty = \sqrt{\langle \sigma_\theta^2 \rangle \langle 1/\sigma_\theta^2 \rangle}
    \label{eq:gaininfty}
\end{equation}
for which Jensen’s inequality ensures that $g_\infty>1$.

\begin{figure}
\begin{center}
\includegraphics[width=\linewidth]{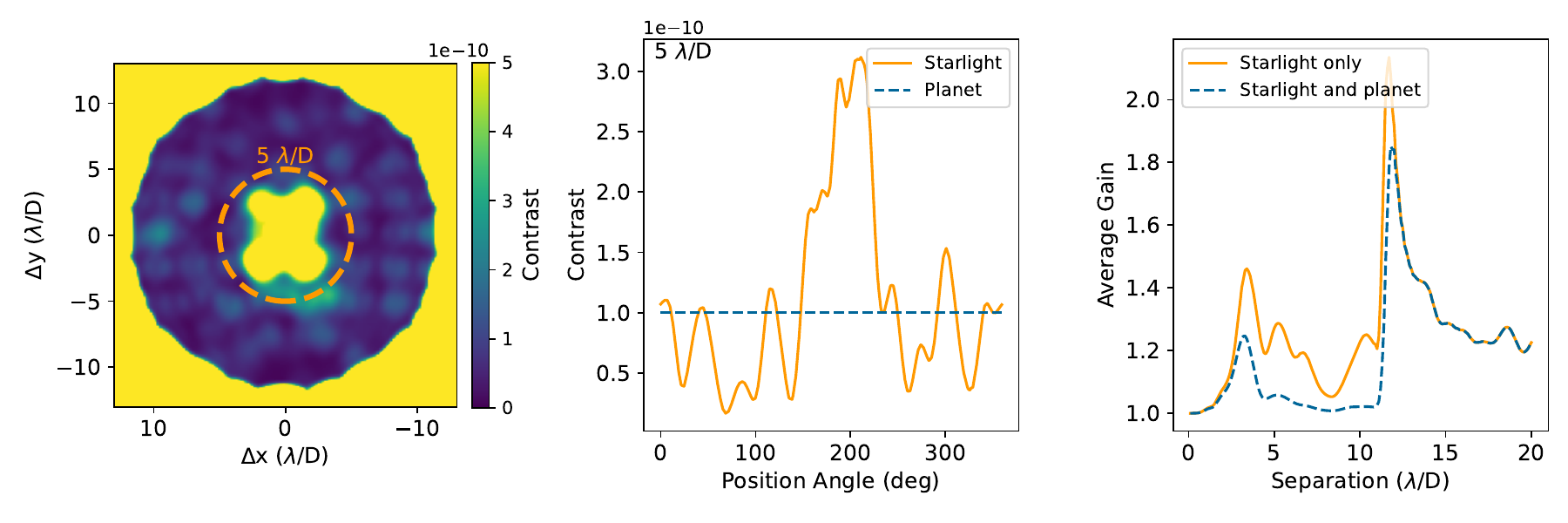}
\end{center}
\caption 
{ \label{fig:psf_avg_gain}
Average sensitivity gain for a simulated PSF. (Left) Image of a simulated dark hole. The PSF was computed using the CDS pipeline\cite{Belikov2024SPIE13092E..66B} and convolved with a $1\lambda/D$-radius aperture. (Middle) Azimuthal profile of the speckle field at $5\lambda/D$ from the star compared to a typical exo-Earth flux ratio of $10^{-10}$. (Right) Average gain $g_\infty$ as a function projected separation assuming a negligible planet flux (Starlight only) or comparable planet-starlight flux (Starlight and planet).} 
\end{figure}

Eq.~\ref{eq:gaininfty} can be computed for any given point spread function and planet flux. 
As an example, we illustrate the average gain as a function of projected separation for a simulated PSF with a dark hole in Fig.~\ref{fig:psf_avg_gain}. The PSF was generated using the CDS pipeline\cite{Belikov2024SPIE13092E..66B} assuming a telescope size of $D=6.513\,$m, wavelength $\lambda=600\,$nm, and a dark hole of 12$\lambda/D$. We also convolved the PSF with an aperture with a radius of $1\lambda/D$ to account for the extent of the planet signal. We note that $5\lambda/D$ corresponds to $1$~au at $10$~pc with these assumptions.
Fig.~\ref{fig:psf_avg_gain} also shows that the gain is negligible when the planet signal is brighter than the background sources in the dark hole.

\section{Case of finite number of rolls}
\label{sec:avggain_fin}

\begin{figure}
\begin{center}
\includegraphics[width=\linewidth]{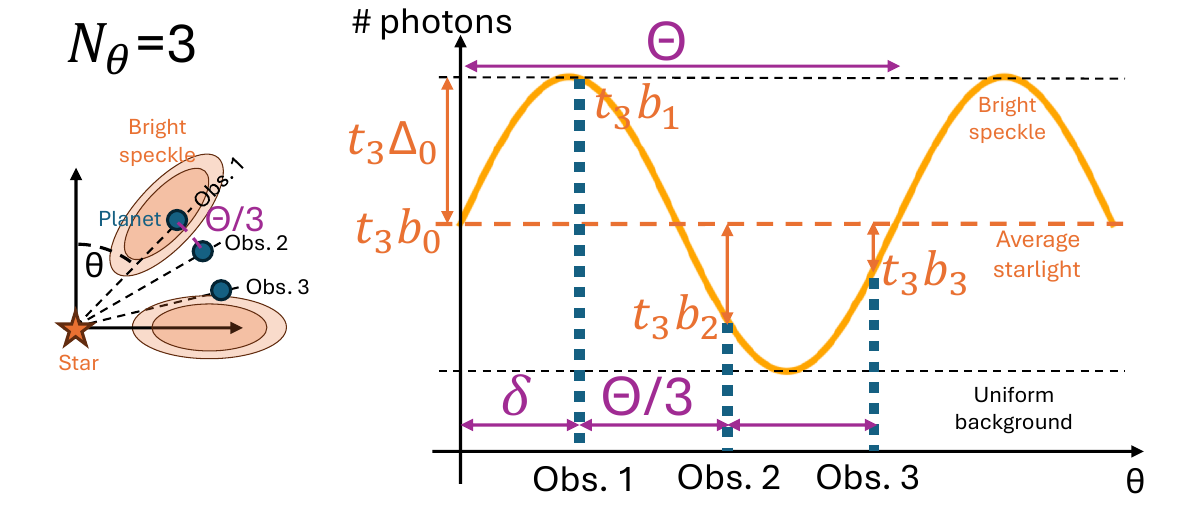}
\end{center}
\caption 
{ \label{fig:rolls}
Illustration of a multiple-roll observations, here assuming $N_\theta=3$, with toy model speckle pattern following a sine wave as a function of azimuth $\theta$. The left panel illustrates the position of the planet relative to speckles in an ADI observation. For any number of observations, the roll angles are defined such as to uniformly sample one period of the sine wave. The right panel illustrates the varying speckle intensity as a function of azimuth. The photon noise sources include a uniform background component, the varying speckle, and the planet.
} 
\end{figure} 

Large number of rolls can be challenging to implement for a space-based observatory due to stability constraints and resulting overheads. As a result, we explore the dependence of the gain to the number of rolls using a toy model as described below. We assume that the speckle intensity varies as the sine of the azimuth angle $\theta$ with a period of $\Theta$. For a given number of rolls $N_\theta$, the angle between each roll is $\Theta/N_\theta$ (Fig.~\ref{fig:rolls}). This prescription ensure that the roll angles cover the full periodicity of the speckles. The phase angle that determines the relative position of the rolls to the speckle pattern is defined as $\delta$ and we have $\delta \in [0, \Theta/N_\theta]$

The roll angle of each observation is given by:
\begin{equation}
     \theta_i = i \frac{\Theta}{N_\theta} + \delta
\end{equation}
and the resulting background flux is  
\begin{equation}
b(\theta_i) = b_0 + \Delta_0 \sin\left( \frac{2 \pi \theta_i}{\Theta} \right)
\end{equation}

The inverse combined variance is given by:

\begin{equation}
\frac{1}{\tilde{\sigma}^2} = \sum_i \frac{1}{t_{N_\theta} \left[ p + b_0 + \Delta_0 \sin\left( \frac{\pi \theta_i}{\Theta} \right), \right] }
\end{equation}

Which leads to the following gain: 
\begin{equation}
    g(\delta) = \frac{p}{\sqrt{p+b_0}} \sqrt{\frac{1}{N_\theta}\sum_{i=1}^{N_\theta} \frac{1}{1 + \frac{\Delta_0/b_0}{p/b_0+1} \sin\left( 2 \pi (\frac{i}{N_\theta} + \frac{ \delta}{\Theta}) \right)} }.
\end{equation}

\begin{figure}
\begin{center}
\begin{tabular}{c}
\includegraphics[height=10cm]{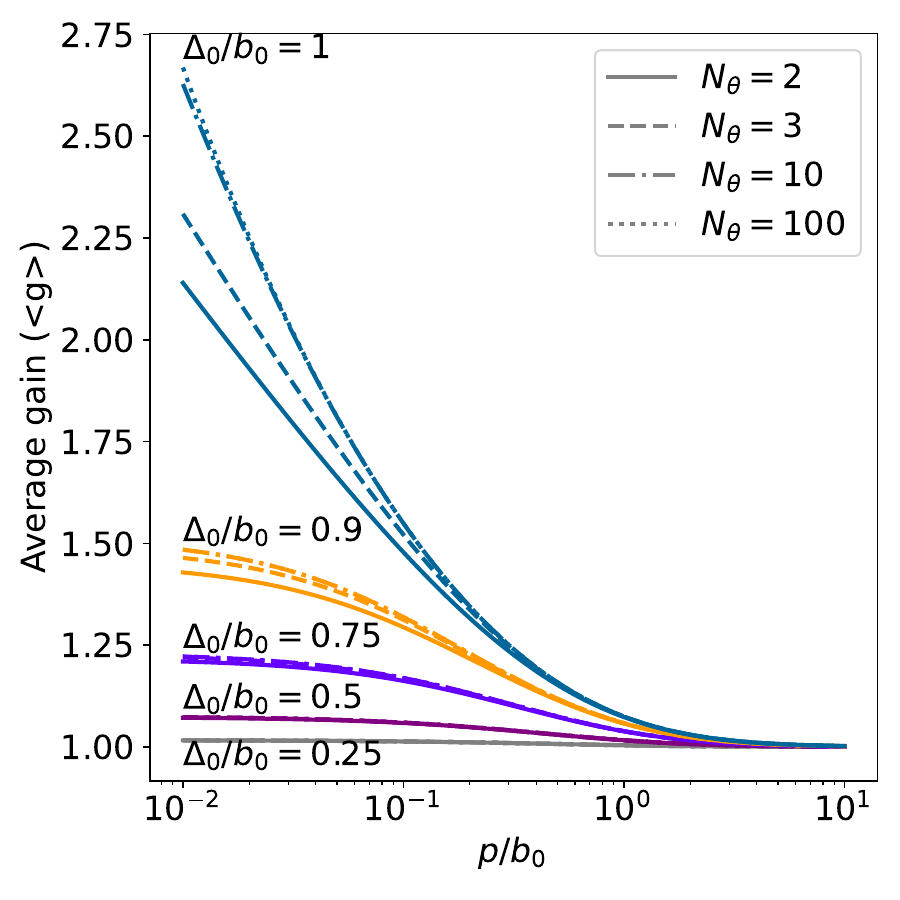}
\end{tabular}
\end{center}
\caption 
{ \label{fig:avg_g}
Average post-processing gain. It is computed as a function of the planet-to-background flux ratio $p/b_0$, the speckle-to-background flux ratio $\Delta_0/b_0$, and the number of roll angles $N_\theta$. The code used to generate this figure is available on Github: \url{https://github.com/jruffio/speckles_snr_gain}.} 
\end{figure}

We then calculate the average gain over $\delta$, which gives:
\begin{equation}
    \left\langle g \right\rangle = \int_0^{\Theta/N_\theta} g(\theta) \, \frac{d\theta}{\Theta/N_\theta}.\label{eq:avggain}
\end{equation}
Finally, we plot the average gain in Fig.~\ref{fig:avg_g}. Once again, the gain is always greater than unity and it is the largest when the speckle photon noise dominates the noise budget. Fig.~\ref{fig:avg_g} also demonstrates that only a few rolls is typically needed to effectively achieve the maximum gain at the limit of infinite rolls. 

\section{Discussion}

The computation of the data processing gain in Eq.~\ref{eq:gaininfty} could be implemented in ETC calculations for HWO to account for the effects of the non-uniformity of any simulated PSF. While this analysis focuses on the average S/N gain, we note that the patchy speckles may also increase the spread of the expected S/N at a given projected separation if the number of rolls is small. This would have an impact on the expected planet completeness for a survey. For example, a small fraction of fainter planets could in fact be detected, while a small fraction of brighter planets would be missed compared to the uniform noise case. This is because the sensitivity is better or worse depending on where the planets lands relative to the bright regions of the speckle pattern. In order to harvest the benefits of the non-uniform speckle pattern, it is important to perform the PSF subtraction on individual images and optimally combining after by leveraging the 2D noise maps. 
The findings in this work are however very general. They apply to any case where two observations are combined with different noise level. This gain would even apply to temporal combination like orbital differential imaging\cite{Males2015SPIE.9605E..18M}. 

Additionally, we postulate that the benefits of non-uniform speckle patterns could be relevant for coronagraph design optimization and speckle-nulling applications as suppressing the starlight uniformly might not be optimal. Indeed, the cost function for dark hole optimization is typically the mean intensity in the region of interest\cite{Mennesson2024JATIS..10c5004M}. Accounting for the average sensitivity gain from non-uniform speckles (Eq.~\ref{eq:gaininfty}), it could be more optimal to optimize $\langle 1/I \rangle$ instead of $\langle I \rangle$ if $I$ is the intensity in the dark hole. While minimizing $\langle I \rangle$ aims at suppressing any bright speckle, maximizing $\langle 1/I \rangle$ on the other hand would allow for bright speckles as long as they also enable comparatively darker regions elsewhere. For example, the non-uniformity of the dark hole is particularly prominent at $\sim3\lambda/D$ in the simulated HWO PSF shown in Figure~\ref{fig:psf_avg_gain}, which is the relevant separation for exo-Earths with typical observatory assumptions at distance $>15\,$pc.

The effect of patchy speckles is a rare example for which neglecting details in the simulation leads to a more pessimistic outcome. 
However, the gain is only significant if the starlight dominates both the planet and the uniform background components of the noise. For example, the exozodi is expected to dominate the noise budget for stars beyond 10 pc, especially at longer wavelengths\cite{Mennesson2024JATIS..10c5004M}.
We find that the gain is less than $\sim$10\% if 1) the average starlight component is less than the other sources of uniform background light combined (e.g. exozodi), or if 2) the planet is brighter than the total average background flux including starlight.
Interestingly, this gain can also be seen as a reduced penalty when increasing the starlight contribution in the noise budget.
We note that a coronagraph-free instrument like JWST/NIRSpec is typically dominated by the patchy starlight. This means that there is significant gains in combining several rolls by increasing the likelihood of a planet to fall in a low intensity region of speckle pattern. 

In this work, we neglected the additional time overheads that would be needed for each roll. The overheads would reduce the effective observing time when increasing the number of rolls. This could be included in future studies as a correction factor on the total exposure time. We note that roll angles are useful for initial planet detection whether or not PSF subtraction is performed using reference stars or the ADI sequence itself. However, for deep atmospheric characterization, it will be important to optimize the timing and roll angle of the observations to ensure that the planet lands in a starlight gap between bright speckles as much possible.

\section{Conclusion}

In this work, we estimated the effects of the patchy nature of the diffracted starlight on ETC calculations compared to the azimuthally-averaged case. Using a weighted mean and assuming perfect PSF subtraction, we showed that combining different observations (e.g., ADI) such that the planet lands either on a bright or a faint spot of the speckle pattern leads to an additional S/N gain beyond $\sqrt{2}$ (Eq. \ref{eq:gain}). We also highlight the average gain in Eq. \ref{eq:gaininfty} when the position of the planet is not known a priori. This gain becomes negligible unless the starlight dominates the noise budget.

\subsection*{Disclosures}
Minor use of ChatGPT was made to assist in coding, figures, and writing latex equations.

\subsection* {Code, Data, and Materials Availability} 
An example Python script is provided on Github\footnote{\url{https://github.com/jruffio/speckles_snr_gain}}.

\subsection* {Acknowledgments}
 We would like to thank the referees for their insightful contributions to this work. In particular, Eq.~\ref{eq:gaininfty} and Section~\ref{sec:avggain_inf} was directly suggested by referee \#2.
 Material presented in this work is supported by the National Aeronautics and Space Administration under Grants/Contracts/Agreements No. 80NSSC25K7300 (J.-B.R.) issued through the Astrophysics Division of the Science Mission Directorate. Any opinions, findings, and conclusions or recommendations expressed in this work are those of the author(s) and do not necessarily reflect the views of the National Aeronautics and Space Administration.


\bibliography{report}   

@INPROCEEDINGS{Currie2023ASPC..534..799C,
       author = {{Currie}, T. and {Biller}, B. and {Lagrange}, A. and {Marois}, C. and {Guyon}, O. and {Nielsen}, E.~L. and {Bonnefoy}, M. and {De Rosa}, R.~J.},
        title = "{Direct Imaging and Spectroscopy of Extrasolar Planets}",
    booktitle = {Protostars and Planets VII},
         year = 2023,
       editor = {{Inutsuka}, S. and {Aikawa}, Y. and {Muto}, T. and {Tomida}, K. and {Tamura}, M.},
       series = {Astronomical Society of the Pacific Conference Series},
       volume = {534},
        month = jul,
        pages = {799},
          doi = {10.48550/arXiv.2205.05696},
archivePrefix = {arXiv},
       eprint = {2205.05696},
 primaryClass = {astro-ph.EP},
       adsurl = {https://ui.adsabs.harvard.edu/abs/2023ASPC..534..799C}
}

@ARTICLE{Marois2006ApJ...641..556M,
       author = {{Marois}, Christian and {Lafreni{\`e}re}, David and {Doyon}, Ren{\'e} and {Macintosh}, Bruce and {Nadeau}, Daniel},
        title = "{Angular Differential Imaging: A Powerful High-Contrast Imaging Technique}",
      journal = {\apj},
         year = 2006,
        month = apr,
       volume = {641},
       number = {1},
        pages = {556-564},
          doi = {10.1086/500401},
archivePrefix = {arXiv},
       eprint = {astro-ph/0512335},
 primaryClass = {astro-ph},
       adsurl = {https://ui.adsabs.harvard.edu/abs/2006ApJ...641..556M}
}

@ARTICLE{Ruffio2024AJ....168...73R,
       author = {{Ruffio}, Jean-Baptiste and {Perrin}, Marshall D. and {Hoch}, Kielan K.~W. and {Kammerer}, Jens and {Konopacky}, Quinn M. and {Pueyo}, Laurent and {Madurowicz}, Alex and {Rickman}, Emily and {Theissen}, Christopher A. and {Agrawal}, Shubh and {Greenbaum}, Alexandra Z. and {Miles}, Brittany E. and {Barman}, Travis S. and {Balmer}, William O. and {Llop-Sayson}, Jorge and {Girard}, Julien H. and {Rebollido}, Isabel and {Soummer}, R{\'e}mi and {Allen}, Natalie H. and {Anderson}, Jay and {Beichman}, Charles A. and {Bellini}, Andrea and {Bryden}, Geoffrey and {Espinoza}, N{\'e}stor and {Glidden}, Ana and {Huang}, Jingcheng and {Lewis}, Nikole K. and {Libralato}, Mattia and {Louie}, Dana R. and {Sohn}, Sangmo Tony and {Seager}, Sara and {van der Marel}, Roeland P. and {Wakeford}, Hannah R. and {Watkins}, Laura L. and {Ygouf}, Marie and {Mountain}, C. Matt},
        title = "{JWST-TST High Contrast: Achieving Direct Spectroscopy of Faint Substellar Companions Next to Bright Stars with the NIRSpec Integral Field Unit}",
      journal = {\aj},
         year = 2024,
        month = aug,
       volume = {168},
       number = {2},
          eid = {73},
        pages = {73},
          doi = {10.3847/1538-3881/ad5281},
archivePrefix = {arXiv},
       eprint = {2310.09902},
 primaryClass = {astro-ph.EP},
       adsurl = {https://ui.adsabs.harvard.edu/abs/2024AJ....168...73R}
}

@ARTICLE{Stark2019JATIS...5b4009S,
       author = {{Stark}, Christopher C. and {Belikov}, Rus and {Bolcar}, Matthew R. and {Cady}, Eric and {Crill}, Brendan P. and {Ertel}, Steve and {Groff}, Tyler and {Hildebrandt}, Sergi and {Krist}, John and {Lisman}, P. Douglas and {Mazoyer}, Johan and {Mennesson}, Bertrand and {Nemati}, Bijan and {Pueyo}, Laurent and {Rauscher}, Bernard J. and {Riggs}, A.~J. and {Ruane}, Garreth and {Shaklan}, Stuart B. and {Sirbu}, Dan and {Soummer}, Remi and {Laurent}, Kathryn St. and {Zimmerman}, Neil},
        title = "{ExoEarth yield landscape for future direct imaging space telescopes}",
      journal = {Journal of Astronomical Telescopes, Instruments, and Systems},
         year = 2019,
        month = apr,
       volume = {5},
          eid = {024009},
        pages = {024009},
          doi = {10.1117/1.JATIS.5.2.024009},
archivePrefix = {arXiv},
       eprint = {1904.11988},
 primaryClass = {astro-ph.EP},
       adsurl = {https://ui.adsabs.harvard.edu/abs/2019JATIS...5b4009S}
}

@ARTICLE{Nemati2020JATIS...6c9002N,
       author = {{Nemati}, Bijan and {Stahl}, H. Philip and {Stahl}, Mark T. and {Ruane}, Garreth J. and {Sheldon}, Leah J.},
        title = "{Method for deriving optical telescope performance specifications for Earth-detecting coronagraphs}",
      journal = {Journal of Astronomical Telescopes, Instruments, and Systems},
         year = 2020,
        month = jul,
       volume = {6},
          eid = {039002},
        pages = {039002},
          doi = {10.1117/1.JATIS.6.3.039002},
       adsurl = {https://ui.adsabs.harvard.edu/abs/2020JATIS...6c9002N}
}

@INPROCEEDINGS{Males2015SPIE.9605E..18M,
       author = {{Males}, Jared R. and {Belikov}, Ruslan and {Bendek}, Eduardo},
        title = "{Orbital Differential Imaging: a new high-contrast post-processing technique for direct imaging of exoplanets}",
    booktitle = {Techniques and Instrumentation for Detection of Exoplanets VII},
         year = 2015,
       editor = {{Shaklan}, Stuart},
       series = {Society of Photo-Optical Instrumentation Engineers (SPIE) Conference Series},
       volume = {9605},
        month = sep,
          eid = {960518},
        pages = {960518},
          doi = {10.1117/12.2188766},
archivePrefix = {arXiv},
       eprint = {1510.02478},
 primaryClass = {astro-ph.IM},
       adsurl = {https://ui.adsabs.harvard.edu/abs/2015SPIE.9605E..18M}
}

@ARTICLE{Mennesson2024JATIS..10c5004M,
       author = {{Mennesson}, Bertrand and {Belikov}, Ruslan and {Por}, Emiel and {Serabyn}, Eugene and {Ruane}, Garreth and {Riggs}, A.~J. Eldorado and {Sirbu}, Dan and {Pueyo}, Laurent and {Soummer}, Remi and {Kasdin}, Jeremy and {Shaklan}, Stuart and {Seo}, Byoung-Joon and {Stark}, Christopher and {Cady}, Eric and {Chen}, Pin and {Crill}, Brendan and {Fogarty}, Kevin and {Greenbaum}, Alexandra and {Guyon}, Olivier and {Juanola-Parramon}, Roser and {Kern}, Brian and {Krist}, John and {Macintosh}, Bruce and {Marx}, David and {Mawet}, Dimitri and {Prada}, Camilo Mejia and {Morgan}, Rhonda and {Nemati}, Bijan and {Pogorelyuk}, Leonid and {Redmond}, Susan and {Seager}, Sara and {Siegler}, Nicholas and {Stapelfeldt}, Karl and {Steiger}, Sarah and {Trauger}, John and {Wallace}, James K. and {Ygouf}, Marie and {Zimmerman}, Neil},
        title = "{Current laboratory performance of starlight suppression systems and potential pathways to desired Habitable Worlds Observatory exoplanet science capabilities}",
      journal = {Journal of Astronomical Telescopes, Instruments, and Systems},
         year = 2024,
        month = jul,
       volume = {10},
          eid = {035004},
        pages = {035004},
          doi = {10.1117/1.JATIS.10.3.035004},
archivePrefix = {arXiv},
       eprint = {2404.18036},
 primaryClass = {astro-ph.IM},
       adsurl = {https://ui.adsabs.harvard.edu/abs/2024JATIS..10c5004M}
}

@ARTICLE{Stark2024JATIS..10c4006S,
       author = {{Stark}, Christopher C. and {Mennesson}, Bertrand and {Bryson}, Steve and {Ford}, Eric B. and {Robinson}, Tyler D. and {Belikov}, Ruslan and {Bolcar}, Matthew R. and {Feinberg}, Lee D. and {Guyon}, Olivier and {Latouf}, Natasha and {Mandell}, Avi M. and {Rauscher}, Bernard J. and {Sirbu}, Dan and {Tuchow}, Noah W.},
        title = "{Paths to robust exoplanet science yield margin for the Habitable Worlds Observatory}",
      journal = {Journal of Astronomical Telescopes, Instruments, and Systems},
         year = 2024,
        month = jul,
       volume = {10},
          eid = {034006},
        pages = {034006},
          doi = {10.1117/1.JATIS.10.3.034006},
archivePrefix = {arXiv},
       eprint = {2405.19418},
 primaryClass = {astro-ph.EP},
       adsurl = {https://ui.adsabs.harvard.edu/abs/2024JATIS..10c4006S}
}

@INPROCEEDINGS{Belikov2024SPIE13092E..66B,
       author = {{Belikov}, Ruslan and {Stark}, Christopher and {Siegler}, Nick and {Por}, Emiel and {Mennesson}, Bertrand and {Redmond}, Susan and {Chen}, Pin and {Fogarty}, Kevin and {Guyon}, Olivier and {Juanola-Parramon}, Roser and {Kasdin}, Jeremy and {Krist}, John and {Mawet}, Dimitri and {Morgan}, Rhonda and {Mejia Prada}, Camilo and {Pueyo}, Laurent and {Ruane}, Garreth and {Sirbu}, Dan and {Stapelfeldt}, Karl and {Trauger}, John and {Zimmerman}, Neil and {Alagao}, Mary Angelie M. and {Carlotti}, Alex and {Chafi}, Jamal and {Doleman}, David and {Gersh-Range}, Jessica and {K{\"o}nig}, Lorenzo and {Leboulleux}, Lucille and {Moody}, Dwight and {Riggs}, A.~J. and {Serabyn}, Eugene and {Snik}, Frans and {Wallace}, Kent},
        title = "{Coronagraph design survey for future exoplanet direct imaging space missions}",
    booktitle = {Space Telescopes and Instrumentation 2024: Optical, Infrared, and Millimeter Wave},
         year = 2024,
       editor = {{Coyle}, Laura E. and {Matsuura}, Shuji and {Perrin}, Marshall D.},
       series = {Society of Photo-Optical Instrumentation Engineers (SPIE) Conference Series},
       volume = {13092},
        month = aug,
          eid = {1309266},
        pages = {1309266},
          doi = {10.1117/12.3020614},
       adsurl = {https://ui.adsabs.harvard.edu/abs/2024SPIE13092E..66B}
}
\bibliographystyle{spiejour}   

\listoffigures

\subsection*{Biographies}
Jean-Baptiste Ruffio is an assistant research scientist studying extra-solar planets in the Astronomy and Astrophysics department at the University of California, San Diego (UCSD). He has been developing statistical and instrumentation techniques to push the frontiers of planet detection and characterization with the largest telescopes in the world, both on the ground and in space. He obtained a master’s degree from the French Aerospace engineering school (ISAE-Supaero) and a Ph.D. in Physics from Stanford University.

Laurent Pueyo is an astronomer at the Space Telescope Science Institute in Baltimore, MD. He has been designing instruments to image planets orbiting other stars and use them to study the properties of distant worlds. He studied at Universit\'{e} d’Orsay and obtained a master's degree from \'{E}cole normale sup\'{e}rieure Cachan followed by a Ph.D. in Mechanical and Aerospace Engineering from Princeton University, NJ.

\end{spacing}
\end{document}